\documentclass[twoside,a4paper,11pt]{sea10}
\usepackage{graphicx}
\usepackage{hyperref}
\usepackage{movie15}
\usepackage{amssymb}
\usepackage{amsmath}
\begin{document}
\pagenumbering{arabic}
\pagestyle{myheadings}
\thispagestyle{empty}
{\flushleft\includegraphics[width=\textwidth,bb=58 650 590 680]{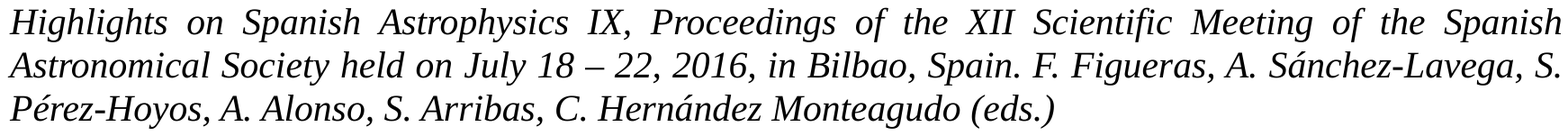}}
\vspace*{0.2cm}
\begin{flushleft}
{\bf {\LARGE
Connecting supernovae with their environments
}\\
\vspace*{1cm}
Llu\'is Galbany$^{1,2}$
}\\
\vspace*{0.5cm}
$^1$Pittsburgh Particle Physics, Astrophysics, and Cosmology Center (PITT PACC). \\
$^2$Physics and Astronomy Department, University of Pittsburgh, Pittsburgh, PA 15260, USA.
\end{flushleft}

\markboth{Connecting SNe with their environments}{L. Galbany}
\thispagestyle{empty}
\vspace*{0.4cm}
\begin{minipage}[l]{0.09\textwidth}
\ 
\end{minipage}
\begin{minipage}[r]{0.9\textwidth}
\vspace{1cm}
\section*{Abstract}{\small
We present MUSE observations of galaxy NGC 7469 from its Science Verification to show how powerful is the combination of high-resolution wide-field integral field spectroscopy with both photometric and spectroscopic observations of supernova (SN) explosions.
Using STARLIGHT and H{\sc ii}explorer, we selected all H{\sc ii} regions of the galaxy and produced 2-dimensional maps of 
%We analyzed all individual spectra in the MUSE datacube with STARLIGHT and produced 2-dimensional maps of continuum-subtracted H$\alpha$ intensity. 
%This was used as an input for H{\sc ii}explorer, a segmentation code that selects all H{\sc ii} regions in the galaxy, in which we calculated 
the H$\alpha$ equivalent width, average luminosity-weighted stellar age, and oxygen abundance. 
%The rotation of the disk in the line-of-sight was estimated from the H$\alpha$ emission shift with respect the galaxy core redshift. 
%
We measured deprojected galactocentric distances for all H{\sc ii} regions, and radial gradients for all above-mentioned parameters.
We positioned the type Ia SN 2008ec in the Branch et al. diagram, and finally discussed the characteristics of the SN parent H{\sc ii} region compared to all other H{\sc ii} regions in the galaxy.
% As an example we show how the the H$\alpha$ equivalent width, the luminosity-weighted stellar age, and the oxygen abundance are distributed across all the {\sc H{\sc ii}} regions in NGC 7469, and estimate gradients of these parameters.
%
In a near future, the AMUSING survey will be able to reproduce this analysis and construct statistical samples to enable the characterization of the progenitors of different supernova types.
\normalsize}
\end{minipage}

%####################################################################
%####################################################################
%####################################################################

\section{Introduction \label{intro}}

Astrophysics has already entered the epoch of big data. 
The current established understanding on several aspects of galaxy formation and evolution now based on a single or low number of observations, will be recalled and challenged by statistical methods applied to the huge amount of data collected by current and new instruments. % and analyzed with statistical methods.
This is the case of present and near-future wide-field and high-resolution Integral Field Spectroscopy (IFS) instruments which have opened new possibilities in several aspects of astrophysics.
Its powerfulness resides on the capability to obtain spectral information at every single position of the instrumental field-of-view with a single exposure ($\sim$100,000 in the case of MUSE).
%, which provide a huge number of spectra in a single exposure (
We are now able to resolve extragalactic H{\sc ii} regions and get the spectral properties of fractions of H{\sc ii} regions.
Old H{\sc ii} region-based relations will need to be recalibrated based not on the whole H{\sc ii} region emission but on fractions with varying ionization conditions.

%In supernova science, the Large Synoptic survey Telescope (LSST) will discover millions of transients that will not be followed-up spectroscopically...
%Similarly, IFS is giving spectra of millions of small regions, which size is seeing-dependent until the advent of IFS + AO.

This new higher spatial precision is also revolutionizing supernova science.
Direct detections of supernova progenitor stars are still rare, and the study of their environment has proven to be a good alternative to put constraints on their characteristics.
IFS is nowadays the most powerful tool to strengthen further these environmental constraints.
After several works that studied the environments of supernovae with IFS \cite{2012A&A...545A..58S,2013A&A...560A..66R, 2013AJ....146...30K,2013AJ....146...31K, 2014A&A...572A..38G, 2016MNRAS.455.4087G,2016A&A...591A..48G} with different spatial and spectral resolutions, the next natural step is to connect those derived environmental parameters with the observed properties of SNe.
This is one of the aims of the All-weather MUse Supernova Integral-field of Nearby Galaxies (AMUSING) survey, which is currently using the MUSE IFS at the Very Large Telescope (VLT) to compile a statistical sample of nearby SN host galaxies, and will definitely be a milestone on supernova environments research.

Here, we present the case of NGC 7469 where SN Ia 2008ec was discovered and followed up, and for which we can connect both environmental and observed properties.

%####################################################################
%####################################################################
%####################################################################

\section{MUSE observations of NGC 7469} \label{sec:data}

NGC 7469 was observed with the MUSE instrument on August 19th 2014 during the the Science Verification of the instrument, under the programme 60.A-9339(A).
Details on the program that proposed the observations and the selection of this galaxy is described in \cite{2016MNRAS.455.4087G} in its Table 1.
Although the quality of the observation is not optimal it has become useful for a series of pilot studies and, in our case, to study SN environments.

\subsection{Analysis}

Reduction and analysis of the datacube is presented in \cite{2016MNRAS.455.4087G} and we refer the reader to that paper for more details.
Below we summarize the common steps, and detail further analysis performed for this work.
%In addition to what we presented there, here we performed the following analysis:
%(1) construct velocity maps from the Halpha velocity shift;
%(2) do kinematic analysis to measure the axis ratio and projection angle, in order to be able to measure deprojected distances to any position on the galaxy disk \cite{2006MNRAS.366..787K};
%(3) measure the deprojected galactocentric distance (GCD) to all {\sc H{\sc ii}} regions found in \cite{2016MNRAS.455.4087G} ;
%(4) produce maps of O/H and HaEW in the H{\sc ii} regions; and 
%(5) produce deprojected maps of these two parameters and measure gradients;

\begin{figure}[!h]
\centering
\includegraphics[trim=0.3cm 0.15cm 0.52cm 0.3cm, clip=true,width=0.495\textwidth]{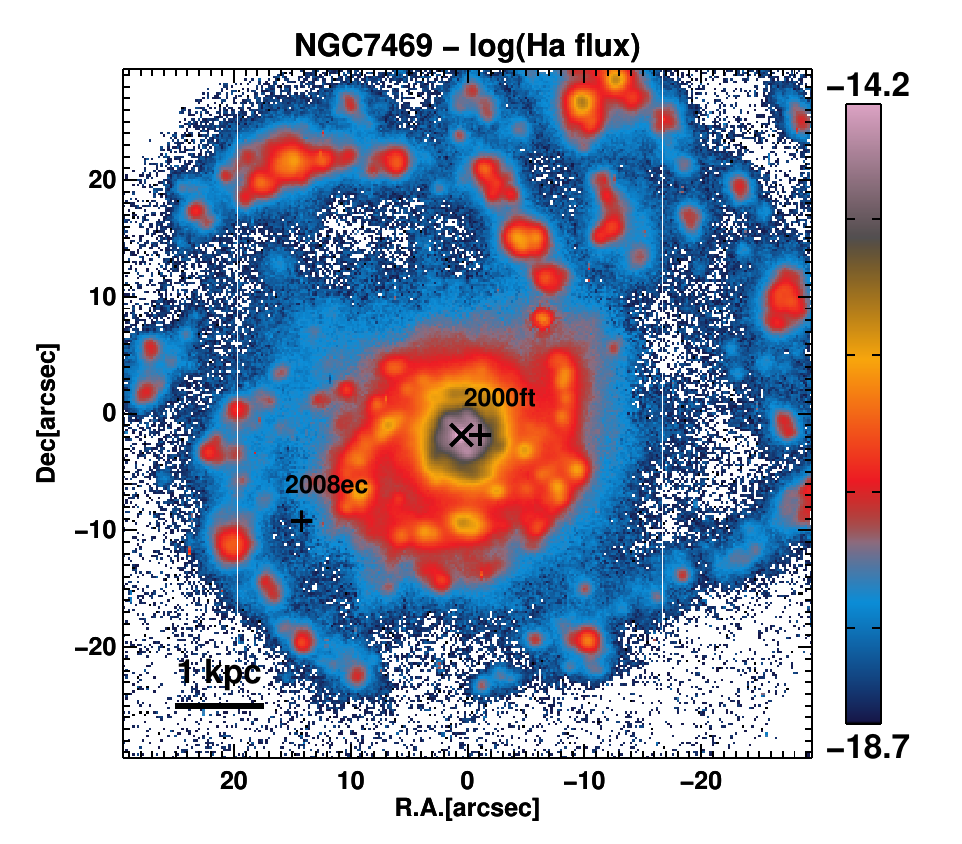}
\includegraphics[trim=0.3cm 0.15cm 0.52cm 0.3cm, clip=true,width=0.495\textwidth]{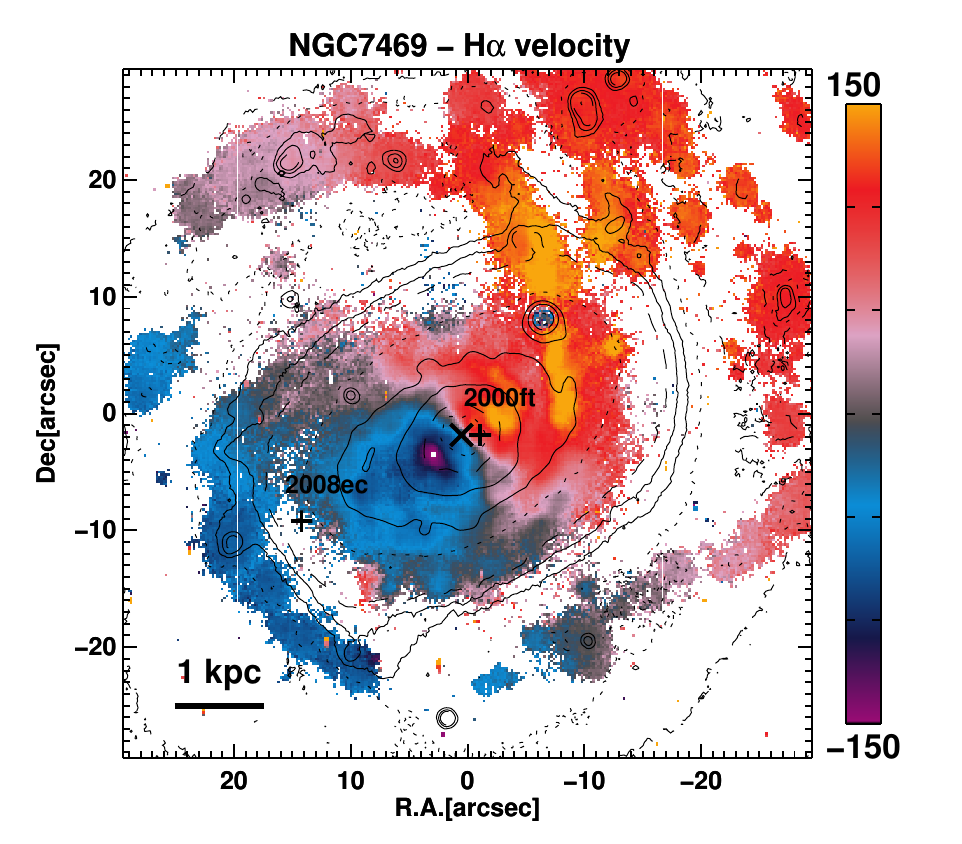}
\includegraphics[trim=0.3cm 0.15cm 0.52cm 0.3cm, clip=true,width=0.32\textwidth]{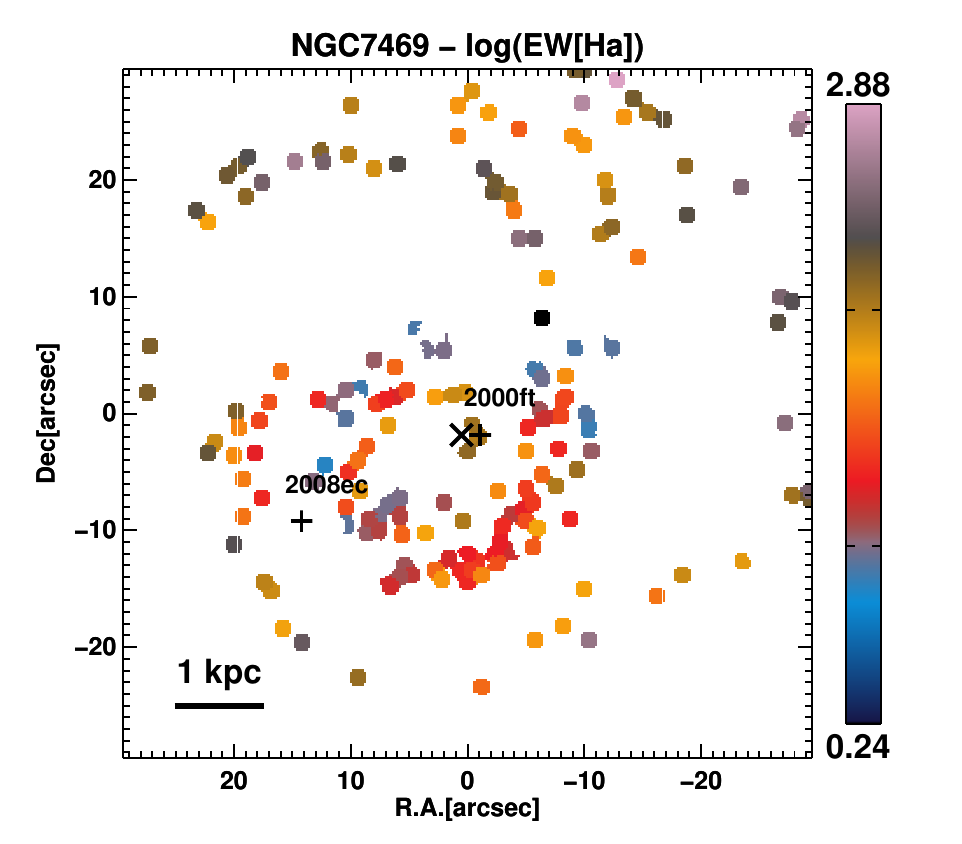}
\includegraphics[trim=0.3cm 0.15cm 0.52cm 0.3cm, clip=true,width=0.32\textwidth]{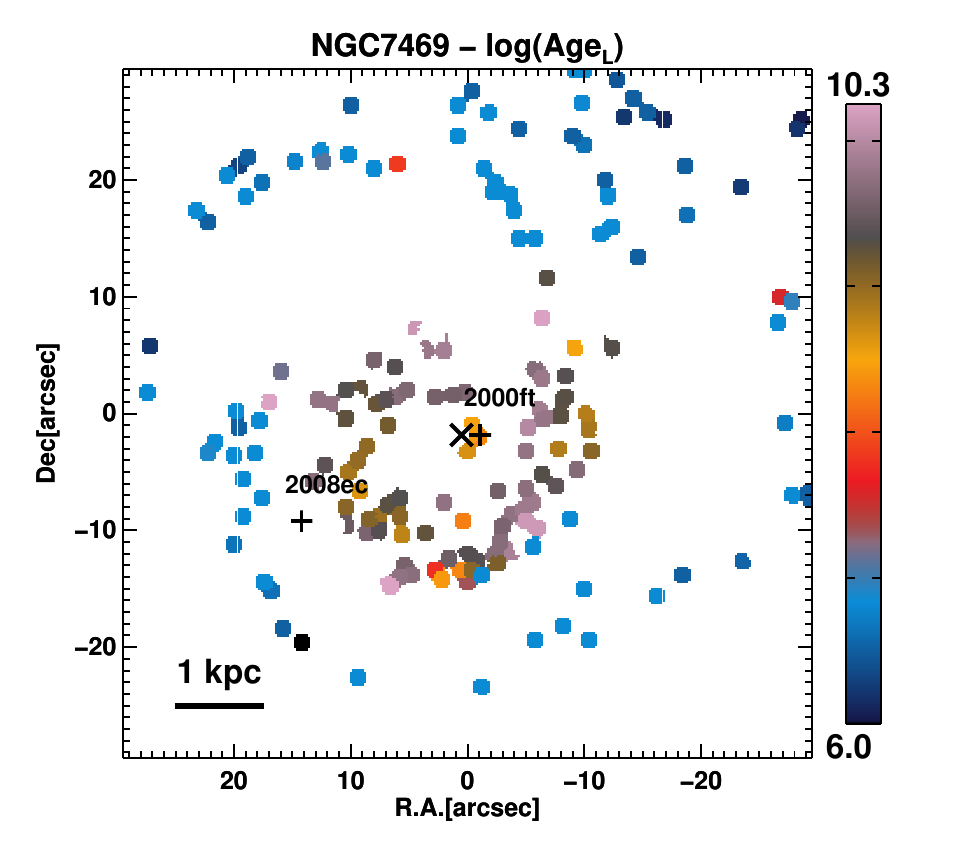}
\includegraphics[trim=0.3cm 0.15cm 0.52cm 0.3cm, clip=true,width=0.32\textwidth]{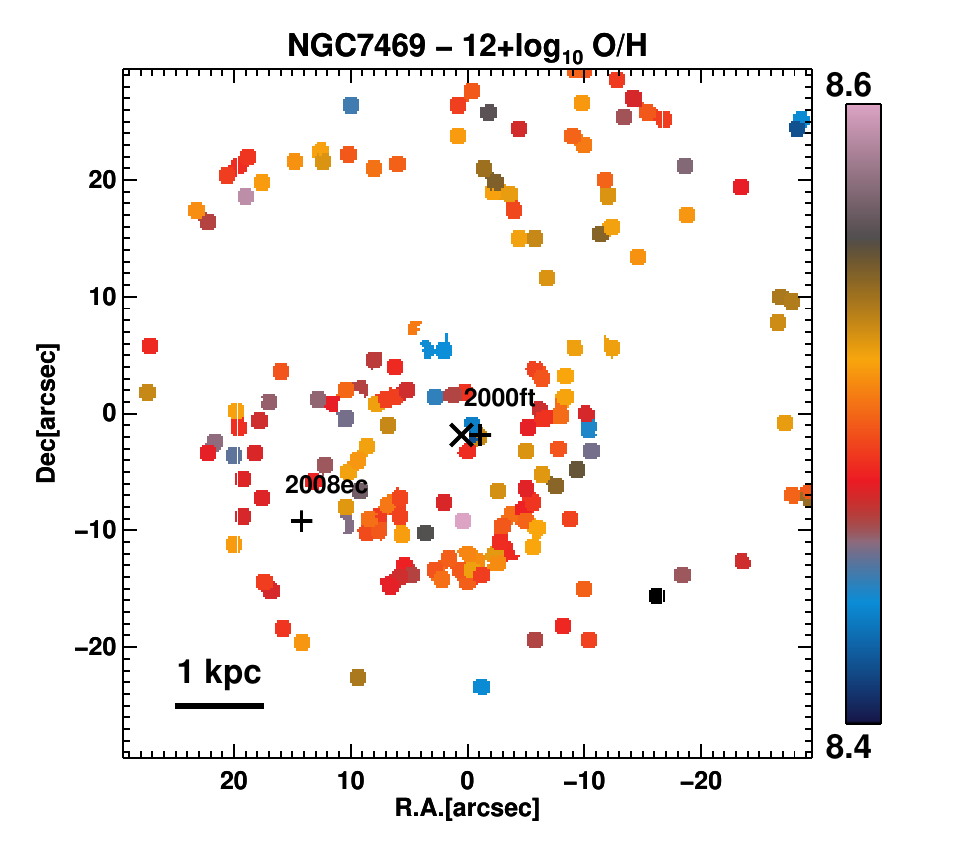}
\includegraphics[trim=0.3cm 0.15cm 0.52cm 0.3cm, clip=true,width=0.32\textwidth]{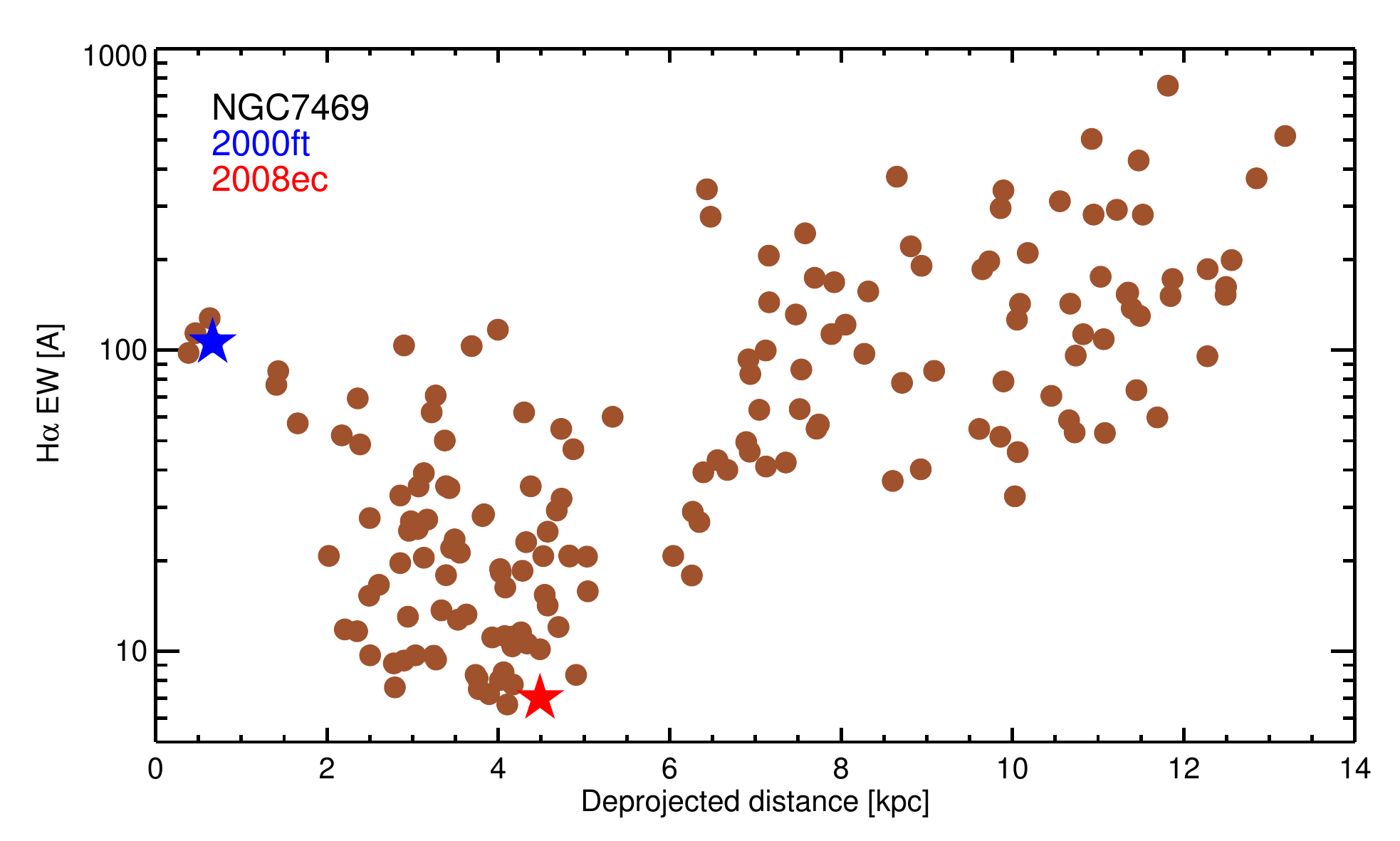}
\includegraphics[trim=0.3cm 0.23cm 0.35cm 0.3cm, clip=true,width=0.32\textwidth]{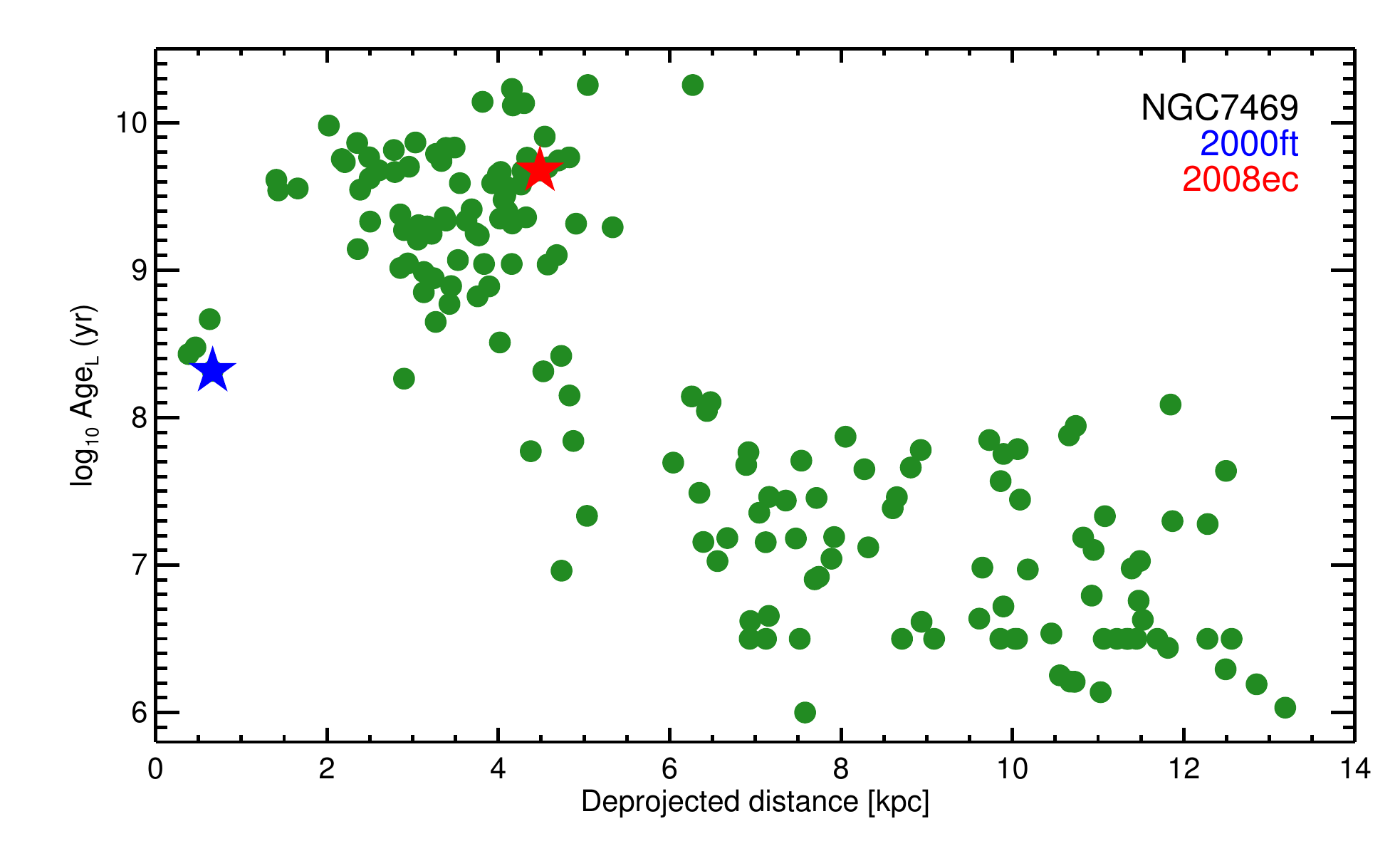}
\includegraphics[trim=0.3cm 0.23cm 0.35cm 0.3cm, clip=true,width=0.32\textwidth]{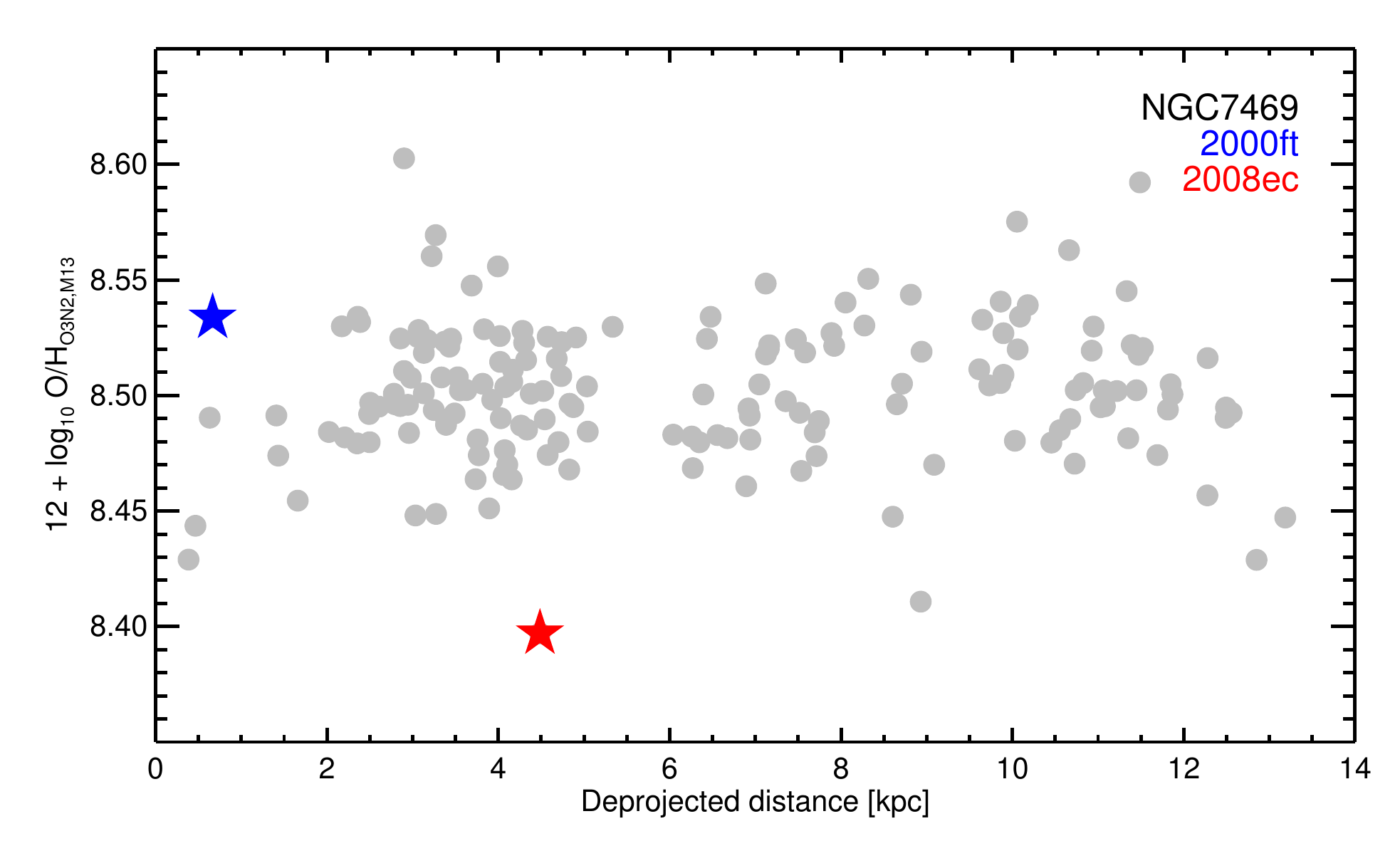}
\caption{Top: H$\alpha$ intensity and velocity maps of NGC 7469 with. We showed with black crosses the galaxy center and the position of SN 20083c and SN 2000ft. Middle: H$\alpha$ equivalent width, average stellar age, and oxygen abundance maps of the H{\sc ii} region segmentation from HIIexplrer, coloured by intensity. Bottom: Radial gradients of these three parameters. We marked with symbols the two SN parent H{\sc ii} regions.}
\label{fig:ngc7469}
\end{figure}

After correcting for Milky Way (MW) dust extinction and shifting the spectra to rest-frame wavelengths, we used {\sc STARLIGHT} \cite{2005MNRAS.358..363C} to fit a combination of single stellar population (SSP) models to all individual spectra ($\sim$100,000) in the datacube. The best combination of SSPs was used to remove the stellar component of the spectra and leave only the pure nebular emission spectra.
Gaussian fits were performed to the Balmer H$\alpha$ emission line to measure their amplitude, width, and shift from the expected rest-frame position.
The shift measured from the H$\alpha$ line has been used to produce velocity maps, which give an idea on the projected galaxy rotation with respect to the line-of-sight.
Kinematic analysis \cite{2006MNRAS.366..787K} were performed to the velocity maps to get estimations on the axis-ratio/ellipticity and projection angle of the galaxy, which were later used to measure deprojected (on-disc) distances from the galaxy core to any position in the galaxy. 
From each individual H$\alpha$ intensity measurement, we created a 2-dimensional map that was given as an input to {\sc H{\sc ii}explorer} \cite{2012A&A...546A...2S}, a package that detects clumps of higher intensity by aggregating adjacent pixels that were then selected as star-forming H{\sc ii} regions.
Following this procedure we were left with 169 H{\sc ii} region spectra.

These 169 spectra were fitted again with STARLIGHT from which we recovered their complete star formation histories. The same procedure described above was repeated to get the pure nebular spectra and fit the strongest emission lines (H$\alpha$, H$\beta$, {\sc Oiii} $\lambda$5007, and {\sc Nii} $\lambda$6583) that were used to estimate the oxygen abundance with the O3N2 calibration proposed by \cite{2013A&A...559A.114M}.
We normalized the spectra using the best SSP fit to the continuum and measured the H$\alpha$ equivalent width, a good proxy for the stellar population age.
Finally, we measured deprojected GCDs to all {\sc H ii} regions to study the radial dependences of these parameters.

In Fig.~\ref{fig:ngc7469}, we show a series of maps resulting from our analysis. The top row shows the H$\alpha$ intensity map and the velocity map measured in each individual spectra, and two SNe discovered in the galaxy with black crosses.
In the mid row, we show the H{\sc ii} region segregation colored in terms of the H$\alpha$EW, the average stellar population age, and the oxygen abundance, respectively.
Below, the radial distributions of these three parameters are shown. On top of them, the stars represent the parent H{\sc ii} regions of SNe discovered in NGC 7469.

\vspace{-0.1cm}
\section{Supernova observed properties}

Two supernovae have been discovered in this galaxy: the type Ia SN 2008ec, and the radio SN 2000ft.
We did a thorough search on the literature and compiled the following information for SN 2008ec:
spectral sequences were presented in \cite{2012MNRAS.425.1789S},
%In addition, we present here eight spectra of SN 2000cb (see section \ref{sec:00cb}) that will be published in Guti\'errez et al. (in prep) in a compilation of SN II spectra from the CATS \citep{2015arXiv151108402G} and CSP (Anderson et al. in prep.) surveys.
%
and multi-band photometry was presented in \cite{2010ApJS..190..418G} and \cite{2010ApJ...721.1627M}.
Since no public data of SN 2000ft is available, we focused our analysis in the former object.

%In  Fig.~\ref{fig:2008ec} we show the multiband light-curve and its 5 spectra...

\begin{figure}[!h]
\centering
\includegraphics[trim=0.8cm 1.0cm 0.4cm 0.9cm, clip=true,width=0.35\textwidth]{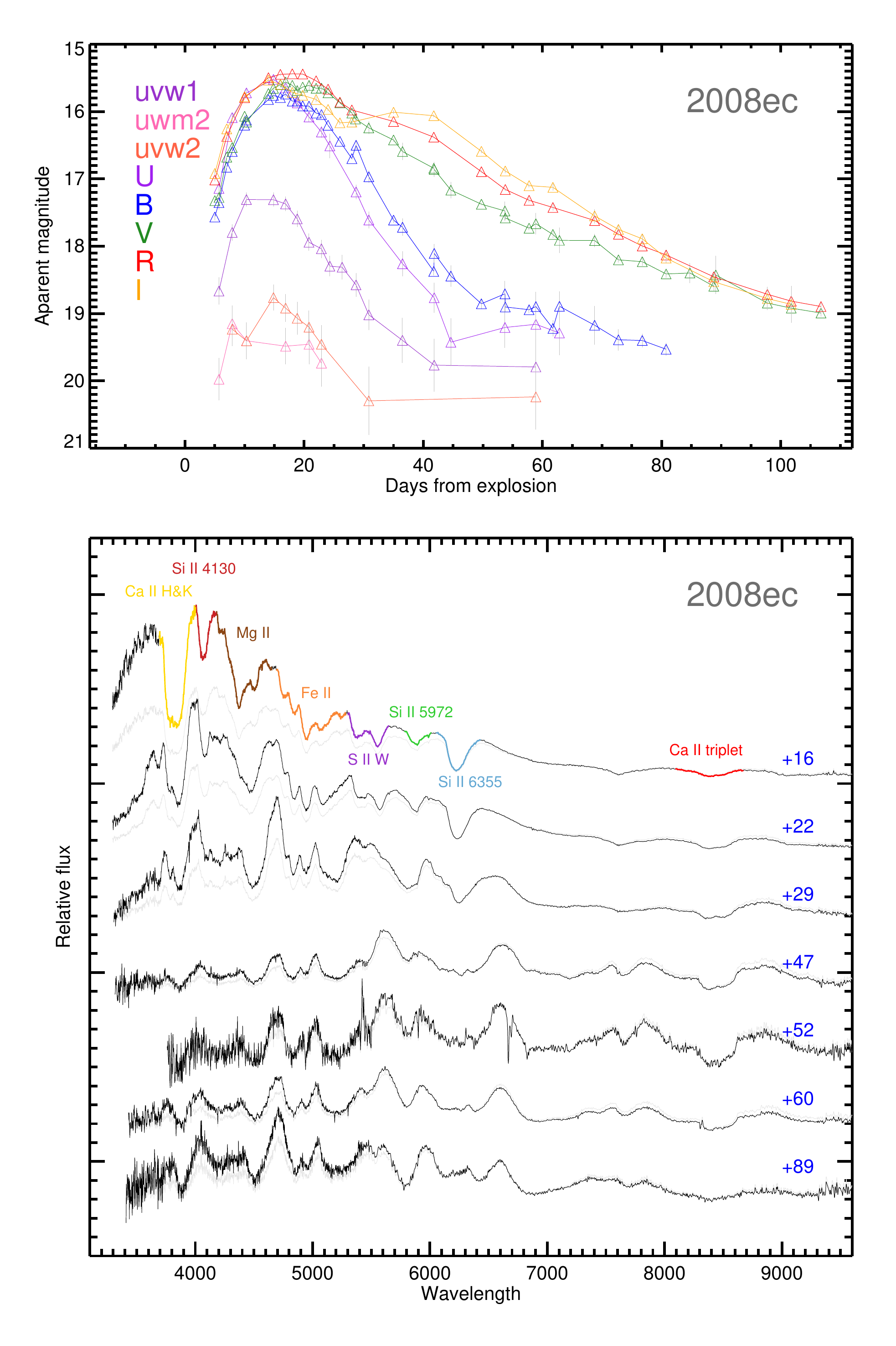}
\includegraphics[trim=0.8cm -1.0cm 0.4cm 0.9cm, clip=true,width=0.55\textwidth]{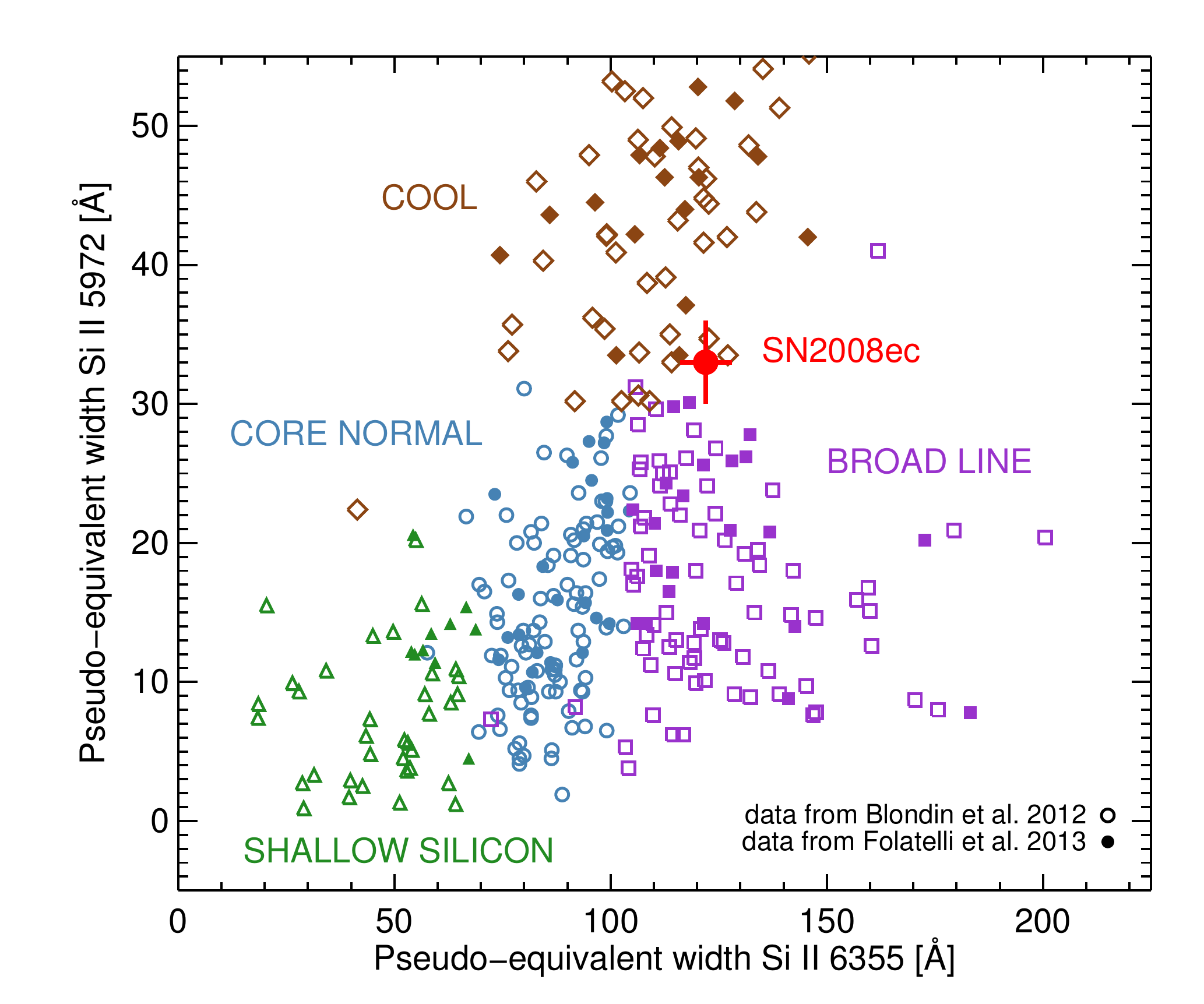}
\caption{Left: SN 2008ec photometry from \protect\cite{2010ApJS..190..418G} and \protect\cite{2010ApJ...721.1627M}, and spectral sequence from \protect\cite{2012MNRAS.425.1789S}. In grey the observed spectra and in black the Milky Way and host galaxy reddening-corrected spectra. Each spectra has been labeled with its epoch relative to explosion. Spectrum at +16 days correspond to the $B$-band maximum light, and the main features have been indicated in colors.
Right: Diagnostic diagram populated with data from the Center for Astrophysics Supernova Program (CfA, open symbols) and from the Carnegie Supernova Project (CSP, filled symbols), with our measurement for SN 2008ec (red) on top.}
% Adapted from Galbany et al. (2016).}
\label{fig:2008ec}
\end{figure}

SN 2008ec was discovered in July 14th 2008 by \cite{2008CBET.1437....1R} and classified as type Ia SN by \cite{2008CBET.1438....1H}.
In Figure \ref{fig:2008ec} we reproduce, left-hand panel, multi-band light-curves referenced to its explosion date (MJD 54657.0) and its spectral sequence where the typical SN Ia features can easily be identified. 
All spectra have been corrected for both Milky Way (MW) extinction using a \cite{1999PASP..111...63F} law and the dust maps of \cite{2011ApJ...737..103S} assuming an $R_V=3.1$,
and for host galaxy reddening with an $R_V=2.21$ and $E(B-V)$=0.24 as found in \cite{2013ApJ...779...38P} by fitting SN multiband light curves. %using NaD $\lambda\lambda$5889,5895 absorptions in high-resolution SN spectra. %+0.53?0.73  %  , Av=0.53+0.13 ?0.16
Spectral epochs with respect the explosion have been also added in labels. Since the rise time in the $B$-band is $\sim$17 days, the earliest spectrum can be considered to have been taken at maximum light.

Although SNe Ia do not show homogeneous brightness at peak, their light-curves can be standardized to reduce the scatter in their peak brightness to around 0.12-0.15 mag.
Several diagnostic diagrams have been proposed (e.g. \cite{2006PASP..118..560B}) to explain such differences, but as discussed in \cite{2016MNRAS.457..525G}, these diagrams give more information about the state of the ejecta than the possible explosion mechanisms progenitor scenarios. In that sense, the proposed subclasses are better described by a continuous sequence than distinct groups, and this is supported by the existence of transitional SN Ia.
To positon SN 2008ec in these diagrams, we measured expansion velocities and pseudo-equivalent widths (pEW) of the two {\sc Si II} $\lambda$5972 and  {\sc Si II} $\lambda$6355 absorption lines in the spectrum corresponding to the epoch of maximum brightness.
We obtained pEW$_{5972}$=34.41$\pm$2.71 \AA~and pEW$_{6355}$=121.78$\pm$1.58 \AA, 
% which is in total agreement with the values reported in \cite{2012MNRAS.425.1819S} (34.4$\pm$2.5 and 118.7$\pm$8.1 \AA), respectively.
Right-hand panel in Figure \ref{fig:2008ec} shows the Branch et al. \cite{2006PASP..118..560B} diagram populated with CfA and CSP data, and our measurements for SN 2008ec, which results to be a transitional event in the boundary between the Cool (CL) and Broad line (BL) SN Ia classes.
CL SNe Ia are a bit fainter than core-normal SNe Ia because of their lower $^{56}$Ni ejecta mass, which mainly controls their temperature and luminosity. % are controlled mainly by the $^{56}$Ni ejecta mass, which is lower for those objects. 
SN 1991bg is the typical object of this class.
On the other hand, BL SNe Ia %show lower Si 5972 pEW and broader Si II 6355 absorptions, which requires high Si II optical depth over a substantial velocity range, since they may 
have thicker silicon layers than the core-normal SN Ia, 
which produce broader {\sc Si II} $\lambda$6355 absorptions given the higher optical depth over a substantial velocity range.
Regarding its expansion velocities, we find v$_{max}^{6355}$=(11.31$\pm$0.16) 10$^3$ km~s$^{-1}$, %, totally in agreement with previous reports (11.31$\pm$0.16 $\times$10$^3$ km~s$^{-1}$, \cite{2014MNRAS.437..338C})
which is within the lower range of the BL SNe Ia, and totally consistent with other CL values.

Branch et al. \cite{2009A&A...500L..17B} presented SALT2 multi-band light-curve fits and reported a $x1$ stretch factor of 1.34 $\pm$ 0.22, which corresponds to $\Delta$m$_{15}$=1.33 $\pm$ 0.19, % according to \cite{2007A&A...466...11G} conversion, 
pointing out that SN 2008ec is an underluminous SN Ia and in absolute agreement with its spectroscopic properties.
%However, \cite{2010ApJ...721.1608B} reported a lower value $\Delta$m$_{15}$=1.08$\pm$0.05 which would indicate that the SN closer to a normal SN Ia.
%\cite{2010ApJ...721.1627M} reported even a lower value (1.06) and showed that its color curves are intermediate to the blue and the red groups of SNe Ia, evolving toward bluer emission.
%\cite{2010ApJS..190..418G} reported a higher value (1.362 $\pm$ 065), and directly measuring the $B$-band magnitude at $\sim$17 and $\sim$32 days past explosion gives 1.31 mag which is closer to the values pointing that it should be an underluminous SN Ia\footnote{Also note that \cite{2011A&A...529L...4C} reported an even lower $x_1$ value, 1.64$\pm$0.18, which would correspond to a $\Delta$m$_{15}\sim$1.39 mag.}.
%\cite{2013MNRAS.436..222M} 0.878$\pm$0.011 with SiFTO

\section{Supernova properties in its galaxy context}

SNe Ia are found in both early- and late-types galaxies which suggests that their progenitors are most probably old stars.
The accepted scenario is a CO white dwarf in a binary system, where the companion can be either another white dwarf (double degenerate, DD) or a younger main sequence star (single degenerate, SD). 
SN Ia in spirals are found to be brighter than those in elliptical galaxies, and it has also been shown that low-stretch SNe Ia occur in regions with older populations \cite{2011ApJ...727..107G}, which may suggest those two different populations in action: 
fainter SNe Ia in early-type galaxies may result from the DD scenario, while brighter SNe Ia in late-type galaxies may come from the SD scenario.

Although NGC 7469 is a late-type galaxy (Sa), SN 2008ec is in an inter-arm region relatively far from any H{\sc ii} region thus indicating that the progenitor star it is not strongly correlated to a region of recent star formation.
The measurements from the spectra of both the closest H{\sc ii} region and at the SN position, show very low H$\alpha$EW values (H{\sc ii}: 6.99$\pm$0.32 \AA, local:  3.14$\pm$0.29 \AA), thus being around the lowest values across the galaxy disk (see lower-left panel in FIg. 1), which correspond to the oldest stellar population ages.
%Age      13.45(0.24) 18.36(2.86)
From the SFH average ages are also relatively old (H{\sc ii}: 5.1$\pm$0.8 Gyr, local:  6.2$\pm$1.0 Gyr), and no younger populations than 1 Gyr are needed in the fit.
Moreover, as seen in Fig. 1, the parent H{\sc ii} region has the lowest oxygen abundance (8.40$\pm$0.05 dex) which is consistent with the value at the SN location (8.41$\pm$0.05 dex).
In addition, both the stretch parameter estimated from their LC fit and the location in the spectroscopic diagram, pointed out that SN 2008ec was relatively faint at peak.
All these indications suggest that their progenitor scenario might not contain young stars, thus  and the most probable is either the DD scenario, or the SD with an older companion.\\

%The local environment can provide more precise information.

%The galaxy has been also observed with IFU in CALIFA \cite{2014A&A...572A..38G}

%\section*{Acknowledgments}
%We would like to thank the MUSE Science Verification team for observing the data analyzed in this work. 
%L.G. was supported in part by the US National Science Foundation under Grant AST-1311862.
%Based on observations made with ESO Telescopes at the La Silla Paranal Observatory under programs 60.A-9339(A).

{\noindent\footnotesize L.G. was supported in part by the US National Science Foundation under Grant AST-1311862.}


\vspace{-0.5cm}
\begin{thebibliography}{}
%\small
%\footnotesize
\scriptsize
\bibitem{2009A&A...500L..17B}{Bailey, S., Aldering, G., Antilogus, P., et al., 2009, A\&A, 500, A17.}
\bibitem{2006PASP..118..560B}{Branch, D., Dang, L. C., Hall, N., et al., 2006, PASP, 118, 560.}
\bibitem{2011A&A...529L...4C}{Chotard, N., Gangler, E., Aldering, G., et al., 2011, A\&A, 529, A4.}
\bibitem{2005MNRAS.358..363C}{Cid Fernandes, R., Mateus, A., Sodr\'e, L., et al., 2005, MNRAS, 358, 363}
\bibitem{1999PASP..111...63F}{Fitzpatrick, E. L., 1999, PASP, 111, 63}
\bibitem{2014A&A...572A..38G}{Galbany, L., Stanishev, V., Mour\~ao, A., et al., 2014, A\&A, 572, A38}
\bibitem{2016A&A...591A..48G}{Galbany, L., Stanishev, V., Mour\~ao, A., et al., 2016, A\&A, 591, A48}
\bibitem{2016MNRAS.455.4087G}{Galbany, L., Anderson, J. P., Rosales-Ortega, F. F., et al., 2016, MNRAS, 457, 525.}
\bibitem{2016MNRAS.457..525G}{Galbany, L., Moreno-Raya, M. E., Ruiz-Lapuente, P., et al., 2016, MNRAS, 457, 525.}
\bibitem{2010ApJS..190..418G}{Ganeshalingam, M., Li, W., Filippenko, A. V., et al., 2010, ApJS, 190, 418.}
\bibitem{2011ApJ...727..107G}{Gonz\'alez-Gait\'an, S., Perrett, K., Sullivan, M., et al., 2011, ApJ, 727, 107.}
\bibitem{2007A&A...466...11G}{Guy, J., Astier, P., Baumont, S., et al., 2007, A\&A, 466, A11.}
\bibitem{2008CBET.1438....1H}{Harutyunyan, A., Benetti, S., Fiorenzano, A., et al., 2008, CBET, 1438}
\bibitem{2006MNRAS.366..787K}{Krajnovi\'c, D., Cappellari, M., de Zeeuw, P. T., et al., 2006, MNRAS, 366, 787}
\bibitem{2013AJ....146...31K}{Kuncarayakti, H., Doi, M., Aldering, G., et al., 2013b, AJ, 146, 31}
\bibitem{2013AJ....146...30K}{Kuncarayakti, H., Doi, M., Aldering, G., et al., 2013a, AJ, 146, 30}
\bibitem{2013A&A...559A.114M}{Marino, R. A., Rosales-Ortega, F. F., S\'anchez, S. F., et al., 2013, A\&A, 559, 114}
\bibitem{2010ApJ...721.1627M}{Milne, P. A., Brown, P. J., Roming, P. W. A., et al., 2010, ApJ, 721, 1627}
\bibitem{2013ApJ...779...38P}{Phillips, M. M., Simon, J. D., Morrell, N., et al., 2013, ApJ, 779, 38}
\bibitem{2008CBET.1437....1R} {Rex, J., \& Filippenko, A. V., 2008, CBET, 1437}
\bibitem{2013A&A...560A..66R}{Rigault, M., Copin, Y., Aldering, G., et al., 2013, A\&A, 560, A66}
\bibitem{2012A&A...546A...2S}{S\'anchez, S. F., Rosales-Ortega, F. F., Marino, R. A., et al., 2012, A\&A, 546, 2}
\bibitem{2011ApJ...737..103S} {Schlafly, E. F., \& Finkbeiner, D. P., 2011, ApJ, 737, 103}
\bibitem{2012MNRAS.425.1789S}{Silverman, J. M., Foley, R. J., Filippenko, A. V., et al., 2012, MNRAS, 425, 1789}
\bibitem{2012A&A...545A..58S}{Stanishev, V., Rodrigues, M., Mour\~ao, A., et al., 2012, A\&A, 545, A58}
\end{thebibliography}
\end{document}